\documentclass{article}

\usepackage[a4paper]{geometry}
\usepackage{graphicx}
\usepackage[colorlinks=true, allcolors=blue]{hyperref}
\usepackage{multirow}
\usepackage{amsmath,amssymb,amsfonts}
\usepackage{amsthm}
\usepackage{mathrsfs}
\usepackage[title]{appendix}
\usepackage{xcolor}
\usepackage{textcomp}
\usepackage{manyfoot}
\usepackage{booktabs}
\usepackage{algorithm}
\usepackage{algorithmicx}
\usepackage{algpseudocode}
\usepackage{listings}
\usepackage{cleveref}
\usepackage{comment}
\usepackage{colortbl}
\usepackage{dsfont}
\usepackage{tikz}
\usetikzlibrary{positioning,decorations.pathreplacing,fit,calc,arrows,tikzmark}
\usepackage{bigstrut}
\usepackage{caption}
\usepackage{subcaption}
\usepackage{tabularx,ragged2e}
\usepackage{longtable,booktabs,threeparttablex}
\usepackage{xltabular}
\usepackage{xtab,afterpage}
\usepackage{orcidlink}

\usepackage[
    style=authoryear,
    natbib=true
]{biblatex}
\usepackage{software-biblatex}
\title{Parameter-Efficient Fine-Tuning of Large Language Models for Unit Test Generation: An Empirical Study}

\author{
  Andr\'e Storhaug \orcidlink{0000-0002-5321-7196}\\
  Norwegian University of Science and Technology, Trondheim, Norway\\
  \texttt{andre.storhaug@ntnu.no}
  \and
  Jingyue Li \orcidlink{0000-0002-7958-391X}\\
  Norwegian University of Science and Technology, Trondheim, Norway \\
  \texttt{jingyue.li@ntnu.no}
}

\date{}

\begin{document}
\maketitle

\begin{abstract}
Parameter-efficient fine-tuning (PEFT) methods, which fine-tune only a subset of model parameters, offer a promising solution by reducing the computational costs of tuning large language models (LLMs) while maintaining their performance. Existing studies have explored using PEFT and LLMs for various code-related tasks and found that the effectiveness of PEFT techniques is task-dependent. The state-of-the-art is limited to using LLMs with full fine-tuning to generate unit tests. The application of PEFT techniques in unit test generation remains underexplored. This paper investigates both full fine-tuning and various PEFT methods, including LoRA, (IA)\textsuperscript{3}, and prompt tuning, across thirteen models of different architectures and sizes. We use well-established benchmark datasets to evaluate their effectiveness in unit test generation and measure syntax correctness, CodeBLEU, pass@1, instruction coverage, branch coverage, and mutation score of the generated tests. Our findings show that LoRA can deliver performance comparable to full fine-tuning for unit test generation in several cases. If training costs are valued, prompt tuning is the most cost-effective approach, particularly for large models. However, the models tuned with full fine-tuning or PEFT may generate fewer executable test cases than the baseline model because they generate more tests calling nonexistent methods or having type mismatches. For the generated ones that are executable, the ones from the tuned models show better test coverage than those from the baseline model.
\end{abstract}

\section{Introduction}\label{sec1}
Training LLMs on extensive code bases like GitHub gives them seemingly magical capabilities. However, despite their proficiency in multiple code-related tasks, LLMs still struggle with the high cost of their training. The traditional approach to handling specialized tasks has been fine-tuning the model for each task. Given that LLMs often have billions, if not trillions, of parameters, traditional fine-tuning, which adjusts all parameters, is extremely computationally expensive and resource-intensive \citep{tufano2021unit}.

Advancements in LLMs have encouraged investigations into various prompting and tuning techniques, e.g., \citet{siddiq2024using, yuan2023manual, schäfer2023empirical}, that do not require full fine-tuning. This challenge has led to the development of multiple techniques for fine-tuning larger models with limited resources, commonly referred to as parameter-efficient fine-tuning (PEFT). The application of various PEFT methods has been explored for code-related tasks, such as code completion, code summarization, and more \citep{Ayupov2022ParameterEfficientFO, Goel2022OnTC, Wang2023PromptTI, weyssow2024exploring, Wang2023OneAF, Shi2023TowardsEF, Liu2023MFTCoderBC, yuan2023evaluating, Liu2023AnES}. Evaluations of the PEFT methods show that they could achieve performance comparable to full fine-tuning while significantly reducing the computational burden by strategically fine-tuning only a chosen subset of parameters for code-related tasks. However, the effectiveness of PEFT techniques on different tasks varies \citet{Ding2023ParameterefficientFO}. 

LLMs have been used to generate unit test cases and assertion statements \citep{Mastropaolo2021StudyingTU, yuan2023evaluating, siddiq2024using, yuan2023manual, schäfer2023empirical, tufano2021unit}. However, to the best of our knowledge, the area of empirical comparisons of PEFT and full fine-tuning for unit test generation remain unexplored. Therefore, in this work, we empirically evaluate the performance of various PEFT methods for unit test generation. We focus on answering the following research questions:
\begin{itemize} 

\item\textbf{RQ1: How well does PEFT perform on unit test generation?}
\item\textbf{RQ2: What is the relation between resource utilization and performance of PEFT in unit test generation?}

\end{itemize}

To answer the research questions, we compared full fine-tuning with three popular PEFT methods: LoRA (Low-Rank Adaptation) \citep{Yu2023LowRankAO}, (IA)\textsuperscript{3} (Infused Adapter by Inhibiting and Amplifying Inner Activations) \citep{Liu2022FewShotPF}, and prompt tuning \citep{Lester2021ThePO}. We conducted the comparisons using thirteen LLMs from four open-source decoder-only LLM families. The LLMs have varying architectures and sizes, ranging from 350 million to 16 billion parameters. We measured and compared the unit test generation performance of full fine-tuning and PEFT methods using well-established benchmark datasets rooted in real-world unit test cases and metrics, including syntactic correctness, CodeBLEU, pass@1, instruction coverage, branch coverage, and mutation score.

Our results show that LoRA is generally the most reliable PEFT method to generate good-quality unit tests and is, in many cases, on par with or even better than full fine-tuning. Regarding cost-effectiveness, prompt tuning provides the best results, particularly for larger models. Prompt tuning can be considered ``low-hanging fruit'', while LoRA should be employed for greater performance. The results also reveal that the generated unit tests from tuned models can contain more compilation errors, which need to be investigated in future studies.  
Our contributions are as follows:
\begin{itemize} 
    \item We conducted the first empirical study to extensively evaluate tuning LLMs using various PEFT methods—LoRA, (IA)\textsuperscript{3}, and prompt tuning—for unit test generation across a wide range of LLM models.
    \item We offered a comprehensive comparison of PEFT methods versus full fine-tuning.
    \item Our findings yield practical guidelines highlighting the potential gains and limitations of the various PEFT methods and full fine-tuning in resource utilization and performance.
\end{itemize}

The rest of the paper is organized as follows. \Cref{chap:peft-methods} introduces the PEFT methods we explore. \Cref{chap:related-work} introduces related work. We explain our experimental design in \Cref{chap:experimental-design}. \Cref{chap:experimental-results} presents the experimental results.  \Cref{chap:discussion} discusses our results and limitations. \Cref{chap:conclusion} concludes the study and proposes future work.

\section{PEFT Methods}
\label{chap:peft-methods}
Parameter-efficient fine-tuning (PEFT) methods provide an alternative way of fine-tuning a pre-trained model by training so-called adapters \citep{Ding2023ParameterefficientFO}. Here, most (if not all) parameters of the pre-trained model are frozen, and only a small number of parameters are trained. These adapters are often orders of magnitude smaller than the full model, making sharing, storing, and loading especially convenient. Moreover, some PEFT methods even facilitate the utilization of mixed-task batches, enabling distinct processing for different examples within a batch \citep{Lester2021ThePO}.
PEFT methods can be classified into three main categories: additive, selective, and reparametrization-based \citep{lialin2023scaling}. Additive methods can be further broken down into adapters \footnote{The term ``adapter'' is also used as the name of the first proposed adapter module for NLP.} and soft prompts.

\subsection{Additive}
This approach focuses on augmenting the pre-trained model with additional parameters or layers. Only the newly added components are trained, keeping the original model weights frozen. Popular techniques within the additive category are (IA)\textsuperscript{3} \citep{Liu2022FewShotPF}, adapters \citep{houlsby2019parameter}, and prompt tuning \citep{Lester2021ThePO}.

\paragraph{Adapters}
Adapter, or ``bottleneck adapter,'' introduced by \citet{houlsby2019parameter}, was arguably the first PEFT method. The class of ``adapters'' introduces small, fully connected networks placed after specific sub-layers within the model. These help the model adapt to the new task without significantly altering the core representation learned during pre-training.

\paragraph{Soft prompts}
These techniques focus on creating the so-called ``soft prompts.'' Prompt tuning, introduced by \citet{Lester2021ThePO}, fine-tuned a subset of the model's input embeddings, allowing the model to adapt to new tasks without altering the model's parameters. The prefix tuning expands the idea to all model layers \citep{li-liang-2021-prefix}. 

\subsection{Selective}
When using selective approaches, only a specific subset of the pre-trained model's parameters is fine-tuned. This approach requires carefully selecting the parameters to be updated, ensuring that they are most relevant to the new task. An example is BitFit by \citet{BenZaken2021BitFitSP}, which only updates the bias parameters.

\subsection{Reparametrization-based}
This category utilizes techniques that reformulate the original model's parameters into a more efficient trainable representation. This makes it possible to work with high-dimensional matrices while simultaneously reducing the number of trainable parameters. LoRA \citep{Yu2023LowRankAO} is a typical reparametrization-based method.

\section{Related Work}
\label{chap:related-work}
This section presents state-of-the-art studies generating unit tests using LLMs and studies applying PEFT approaches to enhance code-related tasks.  

\subsection{Unit test generation using LLMs}
\citet{tufano2021unit} propose an automated unit test case generation tool called \textsc{AthenaTest}. They construct a real-world focal method and developer-written test case dataset named \textsc{Methods2Test} \footnote{ https://github.com/microsoft/methods2test} and train a sequence-to-sequence transformer to generate unit test cases.
\citet{yuan2023manual} propose \textsc{ChatTester}, a ChatGPT-based unit test generation approach, which leverages ChatGPT itself to improve the quality of its generated tests. \citet{schäfer2023empirical} conducted a large-scale empirical evaluation on the effectiveness of LLMs for automated unit test generation without additional training. Similarly to \textsc{ChatTester}, they also use ChatGPT (GPT-3.5-Turbo) and propose a re-prompting approach named \textsc{TestPilot}. \textsc{TestPilot} automatically generates unit tests for all API functions in an npm package, achieving a median statement and branch coverage of 70.2\% and 52.8\% on 1,684 API functions, respectively. \citet{siddiq2024using} use the HumanEval and SF110 datasets to benchmark how well Codex, GPT-3.5-Turbo, and StarCoder can automatically generate unit tests. They evaluate the models based on compilation rates, test correctness, test coverage, and test smells. \citet{alagarsamy2024a3test} propose A3Test, a deep-learning based unit-test generation method that augments test generation with assertion knowledge and verifies naming and signatures. They evaluated the method on 5,278 focal methods from the Defects4J \citep{just2014defects4j} benchmark. The methods by \citet{yuan2023manual, siddiq2024using, schäfer2023empirical, alagarsamy2024a3test} primarily leverage large off-the-shelf LLMs, which can be costly and often lack the flexibility for custom fine-tuning to generate unit tests for specific domains and organizations. Additionally, such prompting-based methods often require significant manual effort to design complex templates and rely on expensive enterprise-grade LLMs like ChatGPT. 

\subsection{PEFT methods on code-related tasks}
Many cutting-edge methods for automatic code generation still involve computationally expensive full fine-tuning. One alternative approach to full fine-tuning is proposed by \citet{steenhoek2023reinforcement}. They use Reinforcement Learning from Static Quality Metrics (RLSQM). As reinforcement learning is a very expensive fine-tuning method, they also see the value of using PEFT methods like LoRA to facilitate the training.

PEFT has emerged as a powerful technique for enhancing code generation with LLMs. Researchers are investigating how to apply PEFT techniques to code-related tasks as a more efficient alternative to full fine-tuning. Results show that the effectiveness of PEFT techniques is task-dependent and not transferable. For example, \citet{Ayupov2022ParameterEfficientFO} evaluated adapters and LoRA on various tasks, finding them effective for code understanding but less so for code generation compared to full fine-tuning. \citet{Wang2023PromptTI} report that prompt tuning of CodeBERT and CodeT5 outperforms full fine-tuning for defect prediction, code summarization, and code translation tasks. \citet{Liu2023AnES} evaluated adapter, LoRA, Prefix tuning, and Mix-And-Match adapter (MAM) \citep{mam} for defect detection, code clone detection, and code summarization. They concluded that ``\textit{PEFT may achieve comparable or higher performance than full fine-tuning in code understanding tasks, but they may exhibit slightly weaker performance in code generation tasks.}'' Similarly, \citet{weyssow2024exploring} evaluated text-to-code generation using LoRA, (IA)\textsuperscript{3}, prompt tuning, and prefix tuning. They compared performance across models of varying sizes (up to 7 billion parameters) against full fine-tuning of smaller models (less than 350 million parameters) and concluded that LoRA is the most effective one.

As existing studies show that PEFT techniques' effectiveness is task-dependent, it is, therefore, necessary to empirically evaluate their effectiveness for unit test generation.  
To our knowledge, there is currently no systematic knowledge of how various PEFT methods behave to tune LLMs for unit test generation. In this study, we focus on autoregressive base code models to minimize the influence of prompt templates on PEFT performance.   

\section{Experimental design}
\label{chap:experimental-design}
The main objective of this study is to gain insight into the performance and cost-effectiveness of fine-tuning LLMs using PEFT for unit test generation.

\subsection{Research Questions}
To answer RQ1 (``\textbf{How well does PEFT perform on unit test generation?}''), we aim to evaluate several different PEFT methods on unit test generation. We also want to compare the PEFT methods with full fine-tuning. Thus, our study covers a wide range of language models for code of various sizes, spanning 350 million to 16 billion parameters. We assess the quality of the generated test cases by different PEFT methods by scoring against real-world human-written reference tests and evaluating the test cases' syntactic correctness, similarity to reference tests, passing rates, instruction and branch coverages, and mutation scores. Further, we examine whether applying PEFT methods for unit test generation adversely affects the pre-trained models' code generation performance, also known as catastrophic forgetting \citep{French1999CatastrophicFI, Luo2023AnES}. Organizations may use the same model to generate both code and unit tests within their integrated development environments (IDEs). This evaluation will help us understand if there are any trade-offs between applying fine-tuning and PEFT methods for unit test generation and maintaining performance on other code-related tasks.

To answer RQ2 (``\textbf{What is the relation between resource utilization and performance of PEFT in unit test generation?}''), we analyze the efficiency of PEFT methods in terms of the computational resources they use and the performance they achieve in unit test generation. 

\subsection{Fine-tuning methods selection}
In line with other empirical studies on PEFT \citep{weyssow2024exploring, Pu2023EmpiricalAO}, we select the most popularly used PEFT methods: LoRA \citep{Yu2023LowRankAO}, (IA)\textsuperscript{3} \citep{Liu2022FewShotPF}, and prompt tuning \citep{Lester2021ThePO}. They are readily available in popular libraries and exemplify most of the primary approaches within PEFT. 

\begin{figure*}[t]
    \centering
    \begin{subfigure}[t]{0.45\textwidth}
        \centering
        \includegraphics[scale=0.75,trim={0 -0.3375in 0 0}]{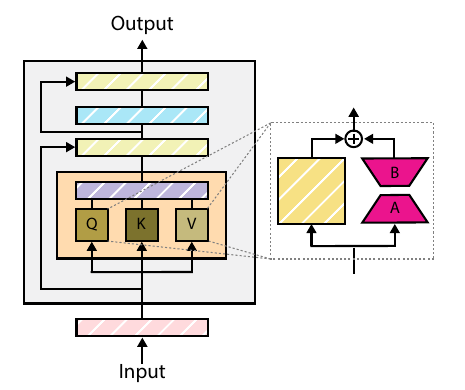}
        \caption{Diagram of LoRA (based on \citet{Yu2023LowRankAO}).}
        \label{fig:lora}
    \end{subfigure}%
    \hfill
    \begin{subfigure}[t]{0.27\textwidth}
        \centering
        \includegraphics[scale=0.75]{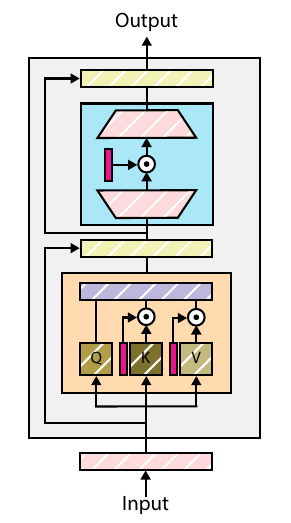}
        \caption{Diagram of (IA)\textsuperscript{3} (based on \citet{Liu2022FewShotPF}).}
        \label{fig:ia3}
    \end{subfigure}
    \hfill
    \begin{subfigure}[t]{0.27\textwidth}
        \centering
        \includegraphics[scale=0.75,trim={0 -0.54in 0 0}]{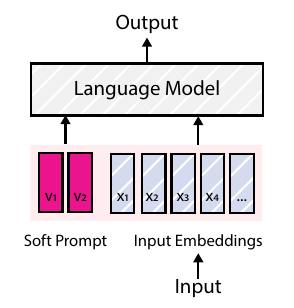}
        \caption{Diagram of prompt tuning (based on \citet{Lester2021ThePO}).}
        \label{fig:prompt-tuning}
    \end{subfigure}
    \caption{PEFT methods architecture illustration diagrams based on the reference implementation of each introductory paper. Striped modules are frozen. Magenta modules are trained.}
\end{figure*}

\begin{itemize}
    \item \textbf{LoRA}: A reparametrization-based method that updates a weight matrix by training a new matrix that is decomposed into a product of two matrices of lower rank: \(\delta W=W_A \cdot W_B\), where \(W_A \in \mathbb{R}^{\text{in} \times r}\) and \( W_B \in \mathbb{R}^{r \times \text{out}}\) \citep{Yu2023LowRankAO}. As demonstrated in \Cref{fig:lora}, the decomposed matrix is then added to the original frozen K and V projection matrices in the attention modules. All other modules are kept frozen.
    \item \textbf{(IA)\textsuperscript{3}}: An additive method that focuses on fine-tuning specific intermediate activations within the model \citep{Liu2022FewShotPF}. As illustrated in \Cref{fig:ia3}, this is done by learning three new vectors: \(l_v, l_k, l_{ff}\); rescaling key, value, and hidden feed-forward activations, respectively.
    \item \textbf{Prompt tuning}: A soft prompts (additive) method that fine-tunes a subset of the model's input embeddings. Exemplified in \Cref{fig:prompt-tuning}, the model input embeddings are prepended with a trainable tensor $P \in \mathbb{R}^{N \times D}$ (known as a ``soft prompt'') \citep{Lester2021ThePO}. Each element of \(P\) is also referred to as a ``virtual token.'' During training, these ``virtual tokens'' are updated while all the rest of the model parameters are kept frozen.
\end{itemize}

\subsection{Models}
\label{sec:models}
We select a diverse set of open-source autoregressive model families trained on code to evaluate the performance and cost-efficiency of PEFT methods. The chosen models span a range of sizes and architectures from the last four years, trained on diverse datasets and training objectives. This offers extensive insights into the interplay between various model capacities and PEFT methods. Further, all models are ensured to support the Java programming language to facilitate a fair comparison of our benchmarking datasets described in \Cref{sec:dataset}.

\begin{itemize}
\item \textbf{Salesforce CodeGen Models}: To democratize program synthesis, Salesforce AI Research developed CodeGen \citep{nijkamp2022codegen}. Pioneering in multi-turn program synthesis, CodeGen presents a variety of sizes, ranging from 350 million to 16 billion parameters. These models have undergone training across a diverse array of both programming and natural languages. The models were trained on ``The pile'' \citep{gao2020pile}, followed by a code subset of Google's BigQuery datasets \citep{GoogleBigQuery}. An updated version CodeGen2 \citep{nijkamp2023codegen2} is also available in sizes 1 billion to 16 billion, trained on ``The Stack'' dataset \citep{Kocetkov2022TheStack} using both causal language modeling (CLM) and file-level span corruption. We select the smallest model of the first generation CodeGen family (\textit{CodeGen-350M-multi}), along with all model sizes of the second generation (1 billion, 3.7 billion, 7 billion, and 16 billion parameters).

\item \textbf{Meta LLaMA Models}: Code Llama \citep{Rozire2023CodeLO}, derived from Llama 2 \citep{Touvron2023Llama2O}, is a specialized model further fine-tuned on the code-specific parts of its pre-training data. With enhanced coding capabilities, it can generate code, interpret natural language prompts, assist in code completion, and aid debugging. It comes in three variants. One is optimized for instructions. Another for Python. We select the base variant \textit{CodeLlama-7B}, as it is designed for general code synthesis and understanding and supports a wide range of popular languages, including Java.

\item \textbf{BigCode StarCoder Models}: This model family includes the 15.5 billion parameters \textit{StarCoderBase} model \citep{li2023starcoder}, along with an updated model generation named StarCoder2 \citep{Lozhkov2024StarCoder2A}, available in 3 billion, 7 billion, and 15 billion parameters. StarCoderBase was trained on 80+ programming languages from ``The Stack'' dataset \citep{Kocetkov2022TheStack}. The updated StarCoder2 is trained on 619 programming languages and other data sources, such as GitHub pull requests, Kaggle notebooks, and code docs. We select all model generations and sizes in this model family. 

\item \textbf{Qwen 2.5 Coder Models}: Qwen 2.5 Coder \citep{hui2024qwen} is the latest code-specific series of Qwen LLMs. The series comes in 0.5, 1.5, 3, 7, 14, 32 billion parameters. The models are built upon the Qwen2.5 architecture and pre-
trained on a vast corpus of over 5.5 trillion tokens. The dataset is sourced from a wide range of public code repositories, as well as large-scale web-crawled data containing code-related texts. We select the 3, 7 and 14 bllion parameter sizes of this model family. 
\end{itemize}

\Cref{tab:model-downloads} shows model statistics for the selected models derived from cfahlgren1/hub-stats hugging face dataset\footnote{\url{https://huggingface.co/datasets/cfahlgren1/hub-stats}} as of 13 November 2025. The chosen models span a representative set of model families released over the past four years, many of which remain highly popular.

\begin{table}[htbp]
    \newcolumntype{Y}{>{\centering\arraybackslash}X}
    \newcolumntype{R}{>{\raggedleft\arraybackslash}X}
    \centering
    \caption{HuggingFace model download statistics from cfahlgren1/hub-stats dataset as of 13 November 2025.}
    \label{tab:model-downloads}
    \small
    \begin{tabularx}{\columnwidth}{lRRY}
    \toprule
    \textbf{Model} & \textbf{Downloads in October 2025} & \textbf{Downloads all time} & \textbf{Created at} \\
    \midrule
    Qwen/Qwen2.5-Coder-14B & 4621 & 185216 & 2024-11-08 \\
    Qwen/Qwen2.5-Coder-3B & 15469 & 213837 & 2024-11-08 \\
    Qwen/Qwen2.5-Coder-7B & 22839 & 453481 & 2024-09-16 \\
    meta-llama/CodeLlama-7b-hf & 2274 & 200519 & 2024-03-13 \\
    bigcode/starcoder2-7b & 17013 & 695181 & 2024-02-20 \\
    bigcode/starcoder2-15b & 5397 & 450456 & 2024-02-20 \\
    bigcode/starcoder2-3b & 192177 & 6168133 & 2023-11-29 \\
    bigcode/starcoderbase & 7185 & 210970 & 2023-05-03 \\
    Salesforce/codegen2-16B\_P & 13 & 37908 & 2023-04-26 \\
    Salesforce/codegen2-7B\_P & 90 & 28639 & 2023-04-26 \\
    Salesforce/codegen2-3\_7B\_P & 42 & 39302 & 2023-04-25 \\
    Salesforce/codegen2-1B\_P & 226 & 77365 & 2023-04-25 \\
    Salesforce/codegen-350M-multi & 4113 & 588888 & 2022-04-11 \\
    \bottomrule
    \end{tabularx}
\end{table}

\subsection{Datasets}
\label{sec:dataset}
\subsubsection{\textsc{Methods2Test\textsubscript{small}}}
To ensure our experiments' validity and practical applicability, we rely on \textsc{Methods2Test} by \citet{tufano2021unit}. It is the largest publicly available corpus of Java unit test cases and corresponding focal methods. This dataset contains 780,000 test cases sourced from 91,000 open-source repositories on GitHub. These real-world test cases represent the complexity and variability inherent in software development, making them an ideal choice for evaluating automated test generation methods. By leveraging \textsc{Methods2Test}, we can assess the performance of the various PEFT methods and models in generating high-quality unit tests that align with real-world coding practices and challenges.

One of the advantages promised by various PEFT methods is the reduction in the amount of data needed for effective training. Thus, we opt to downsize the \textsc{Methods2Test} dataset to enhance efficiency without compromising the representativeness of the data. Specifically, we adopt a strategy wherein only one focal function is selected at random from each class in the dataset. By doing so, we maintain the essence and diversity of the original dataset while significantly reduce its size. Further, the \textsc{Methods2Test} dataset is distributed in deconstructed components. Hence, we assemble it according to the structure seen in \Cref{lst:data-structure} and package it as \textsc{Methods2Test\textsubscript{small}}. The resulting dataset contains 7440 training, 953 validation, and 1017 testing tests. We use this as the primary training dataset.

\subsubsection{\textsc{Methods2Test\textsubscript{runnable}}}
\label{sec:methods2test-runnable}
As the \textsc{Methods2Test} dataset is made up of tests from the real world, executing the tests to measure their passing rates, test coverages, and mutation scores needs significant extra work. This includes retrieving the correct repository version, resolving dependencies, injecting coverage collection tools and mutation testing systems into the existing build system, and finally executing the code in a controlled and safe manner. We construct a runnable subset of \textsc{Methods2Test} by the following steps: 1)  \textbf{Golden commit}: \textsc{Methods2Test} does not provide the repository commits at the time of scraping. Therefore, for each repository, we consider all commits prior to the last commit\footnote{\url{https://github.com/microsoft/methods2test/commit/309ee31cb3752cd8f3ad4b33cfd2b6e5b0ec5029}} on GitHub that updated the Methods2Test dataset (2021-05-18T23:18:51.000Z UTC) in which the focal method and test class files exist as candidate commits. We brute force (begin at the most recent candidate commit and iteratively check older commits)  compare all available dataset data against the repository until we find an exact match---a golden commit. 2) \textbf{Build system}: We target repositories using the Maven build system \citep{apache-maven:3.8.1}, as this is somewhat structured and not as flexible as Gradle. We rely on build file names to determine the build system. Of the 1017 tests in the test split of \textsc{Methods2Test\textsubscript{small}}, 624 of the repositories in \textsc{Methods2Test} are using Maven, 264 are using Gradle, and 129 are using others. 3) \textbf{Runnable}: Each repository in \textsc{Methods2Test}'s test split is then checked for buildability by running Maven and executing the test split of \textsc{Methods2Test\textsubscript{small}}. Due to a lot of old and missing dependencies, only 119 of the 624 Maven repositories are buildable. We collect all tests from \textsc{Methods2Test} with buildable repositories, resulting in 6150 tests. We then validate their runnability by executing every test, obtaining 3712 runnable tests (see \cref{sec:test-execution} for additional test execution details.) 
4) \textbf{Assertion heuristic}: Finally, we filter out any tests that do not contain any of the substrings  ``\texttt{assert}'', ``\texttt{verify}'', or ``\texttt{fail}''. These substrings are common indicators of meaningful test logic in Java-based testing frameworks. Tests lacking these keywords are often empty, incomplete, disabled, or rely on custom test runners and frameworks that bypass standard test assertions. To avoid misleading results—such as treating a test that silently does nothing as correct—we conservatively discard such test cases. This results in a total of 3266 runnable tests. We package it as \textsc{Methods2Test\textsubscript{runnable}}. We use this as the primary testing dataset.
\noindent We package the commit, build system, and execution status as \textsc{Methods2Test\textsubscript{meta}}.

\makeatletter
\newcommand{\gettikzxy}[3]{%
  \tikz@scan@one@point\pgfutil@firstofone#1\relax
  \edef#2{\the\pgf@x}%
  \edef#3{\the\pgf@y}%
}
\makeatother

\newcommand*{\drawBrace}[5][0pt]{%
    \gettikzxy{(pic cs:#3)}{\ax}{\ay}
    \gettikzxy{(pic cs:#4)}{\bx}{\by}
    \draw [decorate,decoration={brace,amplitude=4pt,aspect=#2}]
    (#1,\ay) --(#1,\by) node[pos=#2, right, font=\footnotesize, xshift=4pt] {#5};
    
}%

\definecolor{mygreen}{rgb}{0,0.6,0}
\definecolor{mygray}{rgb}{0.5,0.5,0.5}
\definecolor{mymauve}{rgb}{0.58,0,0.82}

\lstset{ %
  backgroundcolor=\color{white},   
  basicstyle=\footnotesize,        
  breaklines=true,                 
  captionpos=b,                    
  commentstyle=\color{mygreen},    
  escapeinside={\%*}{*)},          
  keywordstyle=\color{blue},       
  stringstyle=\color{mymauve},     
}

\begin{lstlisting}[
    float=!htb,
    name=listing,
    escapechar=!,
    language=java,
    caption={Dataset components: focal method (\textit{fm}), focal class name (\textit{fc}), constructor signatures (\textit{c}), method signatures(\textit{m}), fields (\textit{f}), test case (\textit{t}), test class name (\textit{tc}). Data is assembeld according to \textit{fm}+\textit{fc}+\textit{c}+\textit{m}+\textit{f}+\textit{t}+\textit{tc}. The code is split after the test case signature. Everything above the red line is the \textit{source} input for unit test generation. The code below the red line is the \textit{target} output of the unit test generation.},
    label={lst:data-structure}
    ]
// Focal class
public class Calculator {!\tikzmark{bgnfm}\tikzmark{bgnfc}\tikzmark{bgnc}\tikzmark{bgnm}\tikzmark{bgnf}!

    // Focal method
    public float add(float value) {!\tikzmark{bgnfm}!
        this.previousValue = this.currentValue;
        this.currentValue += value;
        return this.currentValue;
    }!\tikzmark{trmfm}\tikzmark{trmfc}\tikzmark{trmc}!

    // Constructor signatures
    public Calculator();
    public Calculator(float value);!\tikzmark{trmc}!

    // Method signatures
    public float subtract(float value);
    public float multiply(float value);
    public float divide(float value);
    public void reset();!\tikzmark{trmm}!

    // Fields
    public float currentValue;
    public float previousValue;
}!\tikzmark{trmf}!

// Test class
public class CalculatorTest {!\tikzmark{bgntc}!

    // Test case
    @Test!\tikzmark{bgnt}!
    public void testAdd() {
    !\tikzmark{split}!
        Calculator calc = new Calculator();
        assertEquals(5.0, calc.add(5.0f), 0.001);
        assertEquals(7.0, calc.add(2.0f), 0.001);
    }!\tikzmark{trmt}\tikzmark{trmtc}!
}!\begin{tikzpicture}[overlay, remember picture]
    \drawBrace[8.395*1.05]{0.5}{bgnfm}{trmfm}{\textit{fm}};
    \drawBrace[9.45*1.05]{0.72}{bgnfc}{trmfc}{+\textit{fc}};
    \drawBrace[10.4895*1.05]{0.46}{bgnc}{trmc}{+\textit{c}};
    \drawBrace[11.5475*1.05]{0.3}{bgnm}{trmm}{+\textit{m}};
    \drawBrace[12.6*1.05]{0.23}{bgnf}{trmf}{+\textit{f}};
    \drawBrace[8.39*1.05]{0.5}{bgnt}{trmt}{\textit{t}};
    \drawBrace[9.45*1.05]{0.675}{bgntc}{trmtc}{+\textit{tc}};
    %\iftikzmark{line-code-1-start}{\fill[red,overlay] (pic cs:line-code-1-start) circle[radius=4pt];}{\message{No start for 1}}
    %\drawBrace[0pt]{0.2}{bgnfc}{trmfc}{+fc};

    \gettikzxy{(pic cs:split)}{\ax}{\ay}
    \draw[red,thick,xshift=-3pt,yshift=1pt] (0,\ay) -- (\columnwidth,\ay) node[black,pos=.95,below] {\small Target} node[black,pos=.95,above] {\small Source};
\end{tikzpicture}!
\end{lstlisting}

\subsubsection{HumanEval-X}
HumanEval is a benchmarking dataset for code generation, first introduced by \citet{Chen2021EvaluatingLL}. It consists of 164 hand-written programming problems written in Python. For each programming problem, a function signature, docstring, body, and several unit tests are provided. \citet{Zheng2023CodeGeeXAP} extended the dataset to multiple languages, named HumanEval-X. We use the Java subset of the HumanEval-X dataset to evaluate potential catastrophic forgetting effects of fine-tuning.

\subsection{Training}
As the training dataset \textsc{Methods2Test\textsubscript{small}} only includes method signatures for non-focal methods, the unit test generation task can be considered a sequence-to-sequence (seq2seq) problem. Thus, to preserve the code completion capabilities of the decoder-only models described in \Cref{sec:models}, we model the seq2seq task as an autoregressive problem. This is done by only calculating loss based on the \textit{target} code, i.e., the unit test code. We minimize the following autoregressive cross-entropy loss function:
\begin{equation}
\mathcal{L} = - \sum_{t=1}^{T} \mathds{1}_t \cdot \log p(x_t \mid x_{1:t-1}),
\end{equation}
where:
\begin{equation}
\mathds{1}_i = 
\begin{cases}
    1, & \text{if } x_i \neq -100\\
    0, & \text{otherwise}.
\end{cases}
\end{equation}

The model is fed the \textit{source} and \textit{target} concatenated together (see  \Cref{lst:data-structure}). The token values for the \textit{source} are set to -100. The indicator function $\mathds{1}_i$ is used to ignore the \textit{source} in the loss computation.

\begin{table}[htb]
    \centering
    \caption{Model-agnostic hyperparameters for training.}\label{tab:hyperparameters-training}
    \newcolumntype{Y}{>{\centering\arraybackslash}X}
    \small
    \begin{tabularx}{\textwidth}{llY}
        \toprule
        \textbf{Hyperparameter} & \textbf{Method} & \textbf{Value}\\
        \midrule
        \multicolumn{3}{l}{\cellcolor{gray!10}{\textbf{Common}}} \bigstrut \\*
        Optimizer & - &  AdamW \\
        LR schedule & - &  Linear \\
        LR warmup ratio & - &  0.1 \\
        Batch size & - &  1 \\
        Gradient accumulation steps & - &  8 \\
        \# Epochs & - &  3 \\
        Precision & - &  Mixed\\
        \midrule
        \multirow{4}{*}{Learning rate} & Full fine-tuning & 5E-5\\
        & LoRA & 3E-4 \\
        & (IA)\textsuperscript{3} & 3E-4\\
        & Prompt tuning & 3E-3 \\
        \multicolumn{3}{l}{\cellcolor{gray!10}{\textbf{Method specific}}} \bigstrut \\*
        Alpha & LoRA & 32 \\
        Dropout & LoRA & 0.1 \\
        Rank & LoRA & 16 \\
        Virtual tokens & Prompt tuning & 20 \\
       \bottomrule
    \end{tabularx}
\end{table}

\begin{table}[htb]
    \centering
    \small
    \caption{Model-specific hyperparameters for training.}\label{tab:model-specific-hyperparameters-training}
    \begin{tabularx}{\textwidth}{p{3cm}Xll}
        \toprule
        \textbf{Hyperparameter} & \textbf{Method} & \textbf{Model} & \textbf{Value}\\
        \midrule
        \multirow{10}{3cm}{Targeted attention modules} & \multirow{10}{*}{LoRA, (IA)\textsuperscript{3}} & Salesforce/codegen-350M-multi & qkv\_proj \\
        & & Salesforce/codegen2-1B\_P & qkv\_proj \\
        & & Salesforce/codegen2-3\_7B\_P & qkv\_proj \\
        & & Salesforce/codegen2-7B\_P & qkv\_proj \\
        & & Salesforce/codegen2-16B\_P & qkv\_proj \\
        & & meta-llama/CodeLlama-7b-hf & q\_proj, v\_proj \\
        & & bigcode/starcoderbase & c\_attn \\
        & & bigcode/starcoder2-3b & q\_proj, v\_proj \\
        & & bigcode/starcoder2-7b & q\_proj, v\_proj \\
        & & bigcode/starcoder2-15b & q\_proj, v\_proj \\
        & & Qwen/Qwen2.5-Coder-3B & q\_proj,v\_proj \\
        & & Qwen/Qwen2.5-Coder-7B & q\_proj,v\_proj \\
        & & Qwen/Qwen2.5-Coder-14B & q\_proj,v\_proj \\
        \midrule
        \multirow{10}{3cm}{Targeted feedforward modules} & \multirow{10}{*}{(IA)\textsuperscript{3}} & Salesforce/codegen-350M-multi & fc\_out \\
        & & Salesforce/codegen2-1B\_P & fc\_out \\
        & & Salesforce/codegen2-3\_7B\_P & fc\_out \\
        & & Salesforce/codegen2-7B\_P & fc\_out \\
        & & Salesforce/codegen2-16B\_P & fc\_out \\
        & & meta-llama/CodeLlama-7b-hf & down\_proj \\
        & & bigcode/starcoderbase & mlp.c\_proj \\
        & & bigcode/starcoder2-3b & q\_proj, c\_proj \\
        & & bigcode/starcoder2-7b & q\_proj, c\_proj \\
        & & bigcode/starcoder2-15b & q\_proj, c\_proj \\
        & & Qwen/Qwen2.5-Coder-3B & down\_proj \\
        & & Qwen/Qwen2.5-Coder-7B & down\_proj \\
        & & Qwen/Qwen2.5-Coder-14B & down\_proj \\
       \bottomrule
    \end{tabularx}
\end{table}

We use the Transformers \citep{wolf-etal-2020-transformers} and PEFT libraries \citep{peft} from Hugging Face to conduct all training. We use NVIDIA A100 GPUs and leverage DeepSpeed ZeRO-Offload \citep{Ren2021ZeROOffloadDB} where necessary. We use the hyperparameters shown in \Cref{tab:hyperparameters-training} for training. Like other studies, e.g. that of \citet{weyssow2024exploring}, we select hyperparameter values that align with those specified in the original papers of the respective PEFT techniques. \citet{weyssow2024exploring} fine-tuned their models for up to five epochs. Our results show that the losses of most of our fine-tuned models are fairly stable after three epochs\footnote{Available at \url{https://huggingface.co/datasets/andstor/peft-unit-test-generation-experiments}}. Regarding the unspecified hyperparameters, we leave them to library defaults. Following the defaults of the PEFT library, rank decomposition for LoRA is applied to the attention layers. The same attention layers and the default feed-forward modules are targeted for (IA)\textsuperscript{3}. An exhaustive set of selected hyperparameters is provided\footnote{\url{https://huggingface.co/datasets/andstor/peft-unit-test-generation-experiments\#training-hyperparameters}} for reproducing our results. The selected models have context windows ranging from 2048 to 16,384. To make the experiments align with each other, we limit the number of tokens to the smallest window of 2048.

\subsection{Generation}
\label{sec:experimental-design-generation}

\begin{table}[htpb]
    \newcolumntype{L}{>{\raggedright\arraybackslash}X}
    \newcolumntype{Y}{>{\centering\arraybackslash}X}
    \newcolumntype{R}{>{\raggedleft\arraybackslash}X}
    \centering
    \small
    \caption{Hyperparameters for generation.}
    \label{tab:hyperparameters-generation}
    \begin{tabularx}{\textwidth}{lR}
    \toprule
    \textbf{Hyperparameters} & \textbf{Value} \\
    \midrule
    Do sample                         & False \\
    Temperature                           & 0 \\
    Top p                                & 0 \\
    Frequency penalty                     & 0 \\
    Max length                         & 2048 \\
    \bottomrule
    \end{tabularx}
\end{table}

For generating unit tests with the \textsc{Methods2Test\textsubscript{runnable}} dataset, we provide the \textit{source} part of the data (exemplified in  \Cref{lst:data-structure}) as input to the models to generate unit test cases. In order to facilitate reliable execution during later evaluation stages, the source also includes the test method function definition. For the evaluation of catastrophic forgetting, we use the default evaluation protocol for the HumanEval-X dataset. This includes generating the body of a focal method, given an function signature and docstring as the \textit{source} input.

As we use a context window of 2048, we filter out any samples where the source + target overflows. We implement a stopping strategy based on matching braces to generate well-formed functions. That is, functions that have matching start and end braces. For incomplete unit tests, we follow the protocol from \citet{siddiq2024using} where we iteratively delete lines (from bottom to top), add one to two curly brackets, and check if the syntax is valid using an ANTLR v4-based \citep{antlr4} Java parser \citep{java-antlr4}. We only evaluate samples that are well-formed. We use greedy decoding without any sampling or frequency penalty.
\Cref{tab:hyperparameters-generation} shows the hyperparameter details used for generation. 

\subsection{Metrics}
\subsubsection{Syntactic Correctness}
Following \citet{yuan2023manual}, we measure the syntactic correctness of the generated unit tests, ensuring that they conform to the grammar of the Java programming language. After the fixing described in \Cref{sec:experimental-design-generation}, we employ the same parser as described in \Cref{sec:experimental-design-generation}. This involves parsing the generated tests and verifying that they are free of syntax errors. We then report the percentage of unit tests with valid syntax for each model.

\subsubsection{Similarity}
BLEU score by \citet{Papineni2002BleuAM} is a very common metric used to evaluate natural language machine translation. It measures the similarity of the machine-translated text to the reference translations. However, \citet{Ren2020CodeBLEUAM} recognized that the BLEU score neglects important syntactic and semantic features of programming languages. To remedy this, they developed CodeBLEU \citep{Ren2020CodeBLEUAM}. Based on n-gram matching used in BLEU, they further augment the scoring by leveraging abstract syntax trees (AST) and code semantics via data flow. To evaluate the primary performance of the various PEFT methods, we use CodeBLEU to measure the similarity between the generated unit test code and the ground truth. In our experiments, we use the tokenizer\_13a \footnote{\url{https://github.com/mjpost/sacrebleu/blob/master/sacrebleu/tokenizers/tokenizer_13a.py}} from SacreBLEU by \citet{bojar2018call}, equivalent to the standard mteval-v13a tokenizer used by the Conference on Machine Translation (WMT).

\subsubsection{Passing Rate} \label{sec:pass-rate}
We use pass@1, a special case of the pass@k metric introduced by \citet{Chen2021EvaluatingLL}, to assess the correctness of generated outputs. In our setting, given a single focal method and one corresponding generated test case, pass@1 measures the probability that the test passes when executed. This reflects a realistic deployment scenario where there is only one chance to provide the user with a useful generated test case. Since our generation does not involve sampling, this also ensures reproducibility. 

\subsubsection{Coverage}
To evaluate the effectiveness of the tests generated using \textsc{Methods2Test\textsubscript{runnable}}, we calculate the coverage of the tests that pass. Specifically, we measure instruction coverage and branch coverage. These metrics provide a good indication of how thoroughly the test cases exercise the method under test.

\subsubsection{Mutation Score}
We evaluating the semantic strength of the generated tests by conducting mutation testing of the . Mutation testing evaluates the fault-detection capability of test cases by introducing small artificial defects, so-called mutants, into the program under test and checking whether the generated tests can detect them. A mutant is considered killed if the test fails when the mutation is present. We compute the mutation score as the number of mutants killed divided by the total number of mutants, following standard practice in software testing research \citep{jia2011an}.

\subsection{Test Execution}
\label{sec:test-execution}
To safely execute untrusted code, we utilize Docker containers that isolate the test environment. Java tests from our datasets are executed using the Apache Maven build system~\citep{apache-maven:3.8.1}. For the \textsc{Methods2Test\textsubscript{runnable}} dataset, the Maven configuration is provided by cloning the test's repository. For HumanEval-X, we use a hardcoded setup of Maven, and write the test and focal method to files. To measure code coverage, we employ the JaCoCo Java Code Coverage Library~\citep{jacoco:0.8.3}. For conducting mutation testing, we use the JVM mutation testing system Pitest \citet{pitest:1.16.0}. Pitest inserts mutants into jvm by re-writing the class after it has loaded. We use the following default mutators: ``Conditionals Boundary'', ``Increments'', ``Invert Negatives'', ``Mat'', ``Negate Conditionals'', ``Void Method Calls'', ``Empty returns'', ``False Returns'', ``True returns'', ``Null returns'', and ``Primitive return''.

Prior to execution, we replace the original implementation with the generated content. In the case of \textsc{Methods2Test\textsubscript{runnable}}, this involves replacing the test method. For HumanEval-X, it involves replacing the focal method. We then instruct Maven to execute the single corresponding unit test method and rely on Maven to retrieve any dependencies and build required modules.

Test execution results are obtained from the XML output produced by the standard Maven Surefire Plugin \citep{mavensurefireplugin}. A test run may conclude in several ways. It may ``succeed'' or ``fail'' based on assertion outcomes. It may also result in an error. Errors typically arise from unexpected runtime issues, such as unhandled exceptions during execution---an ``interruption''. Beyond execution-specific issues, the environment can also lead to problems. These include ``compilation errors''---often due to missing references, type mismatches, or unfulfilled dependencies. Although our datasets generally exclude unbuildable codebases (see \cref{sec:methods2test-runnable}), compilation errors can still occur with generated content. Other environmental exceptions may arise during container setup, file I/O, or subprocess invocation, causing some tests to fail to run. 

\section{Experimental Results}
\label{chap:experimental-results}
This section presents the detailed results of each research question.

\subsection{RQ1: Tuning method performance}
We begin by describing the performance results for each evaluation metric on the \textsc{Methods2Test\textsubscript{runnable}} dataset, as shown in \Cref{tab:eval-summary}(b). We then explain the results to test the catastrophic forgetting
effects on code generation (shown in \Cref{tab:eval-summary}(c)) using the HumanEval-X dataset.

\begin{ThreePartTable}
    \newcolumntype{Y}{>{\centering\arraybackslash}X}
    \newcolumntype{R}{>{\raggedright\arraybackslash}X}
    \newcolumntype{L}{>{\raggedleft\arraybackslash}X}
    \centering
    \renewcommand{\arraystretch}{1.25}
    \footnotesize
    \begin{TableNotes}[flushleft, para]\small
      \item \textbf{Bold}: best-performing training method per model. (Parentheses): decreased performance compared to baseline. \colorbox{red!10}{Red}: $<$ 50\% syntactical valid samples.
    \end{TableNotes}
    \begin{xltabular}{\textwidth}{lr!{\color{white}\hspace{.5em}}YYYYYY!{\color{white}\hspace{1em}}YYY}
    \caption{Evaluation metrics experiment results using different tuning methods across various models.}\label{tab:eval-summary}\\
        \multicolumn{2}{c}{\normalsize\textbf{(a)}} & \multicolumn{6}{c}{\normalsize\textbf{(b)}} & \multicolumn{3}{c}{\normalsize\textbf{(c)}}\\[.5em]
        \cmidrule(lr){1-2}\cmidrule(lr){3-8}\cmidrule(lr){9-11}
        \multirow{2}{*}{\textbf{Method}} & \multirow{2}{*}{\parbox[t]{1cm}{\centering \textbf{Trainable\\params}}} & \multicolumn{6}{c}{\textbf{\textsc{Methods2Test\textsubscript{runnable}}}} & \multicolumn{3}{c}{\textbf{\textsc{HumanEval-X\textsubscript{java}}}}\\
        \cmidrule(lr){3-8}\cmidrule(lr){9-11}
        & & \rotatebox[origin=l]{90}{Valid syntax} & \rotatebox[origin=l]{90}{CodeBLEU} & \rotatebox[origin=l]{90}{pass@1} & \rotatebox[origin=l]{90}{Instr. Cov.} & \rotatebox[origin=l]{90}{Branch Cov.} & \rotatebox[origin=l]{90}{Mutation Score} & \rotatebox[origin=l]{90}{Valid syntax} & \rotatebox[origin=l]{90}{CodeBLEU} & \rotatebox[origin=l]{90}{pass@1}\\
        \hline
        \endfirsthead
            \caption{(continued) Evaluation metrics experiment results using different tuning methods across various models.}\\
        \multicolumn{2}{c}{\normalsize\textbf{(a)}} & \multicolumn{6}{c}{\normalsize\textbf{(b)}} & \multicolumn{3}{c}{\normalsize\textbf{(c)}}\\[.5em]
        \cmidrule(lr){1-2}\cmidrule(lr){3-8}\cmidrule(lr){9-11}
        \multirow{2}{*}{\textbf{Method}} & \multirow{2}{*}{\parbox[t]{1cm}{\centering \textbf{Trainable\\params}}} & \multicolumn{6}{c}{\textbf{\textsc{Methods2Test\textsubscript{runnable}}}} & \multicolumn{3}{c}{\textbf{\textsc{HumanEval-X\textsubscript{java}}}}\\
        \cmidrule(lr){3-8}\cmidrule(lr){9-11}
        & & \rotatebox[origin=l]{90}{Valid syntax} & \rotatebox[origin=l]{90}{CodeBLEU} & \rotatebox[origin=l]{90}{pass@1} & \rotatebox[origin=l]{90}{Instr. Cov.} & \rotatebox[origin=l]{90}{Branch Cov.} & \rotatebox[origin=l]{90}{Mutation Score} & \rotatebox[origin=l]{90}{Valid syntax} & \rotatebox[origin=l]{90}{CodeBLEU} & \rotatebox[origin=l]{90}{pass@1}\\
        \hline
        \endhead
        \bottomrule
        \multicolumn{11}{r}{to be continued on the next page}
        \endfoot
        \bottomrule
        \insertTableNotes
        \endlastfoot
        \multicolumn{11}{l}{\cellcolor{gray!10}{\textbf{CodeGen-350M-multi}}} \bigstrut \\*
        None & 0 & 0.96 & 0.17 & 0.09 & \textbf{0.52} & 0.12 & 0.23 & \textbf{1.0} & 0.28 & \textbf{0.03} \\*
        Fine-tuning & 356,712,448 & \textbf{0.98} & \textbf{0.22} & 0.14 & (0.48) & \textbf{0.16} & \textbf{0.83} & \textbf{1.0} & (0.24) & (0.02) \\*
        LoRA & 1,310,720 & 0.96 & 0.19 & \textbf{0.15} & (0.48) & 0.15 & 0.41 & \textbf{1.0} & \textbf{0.29} & (0.02) \\*
        (IA)\textsuperscript{3} & 143,360 & (0.95) & 0.17 & 0.13 & (0.5) & 0.13 & 0.45 & \textbf{1.0} & 0.28 & \textbf{0.03} \\*
        Prompt tuning & 20,480 & 0.96 & 0.17 & 0.12 & (0.43) & 0.13 & 0.44 & \textbf{1.0} & (0.25) & (0.02) \\

        \multicolumn{11}{l}{\cellcolor{gray!10}{\textbf{CodeGen2-1B}}} \bigstrut \\*
        None & 0 & \cellcolor{red!10}{0.0} & \cellcolor{red!10}{0.0} & \cellcolor{red!10}{0.0} & \cellcolor{red!10}{0.0} & \cellcolor{red!10}{0.0} & \cellcolor{red!10}{0.0} & \cellcolor{red!10}{0.0} & \cellcolor{red!10}{0.0} & \cellcolor{red!10}{\textbf{0.0}} \\*
        Fine-tuning & 1,015,306,240 & \textbf{0.75} & 0.1 & 0.04 & \textbf{0.31} & \textbf{0.09} & 0.11 & \cellcolor{red!10}{0.05} & \cellcolor{red!10}{0.01} & \cellcolor{red!10}{\textbf{0.0}} \\*
        LoRA & 2,097,152 & \cellcolor{red!10}{0.38} & \cellcolor{red!10}{0.03} & \cellcolor{red!10}{\textbf{0.12}} & \cellcolor{red!10}{0.01} & \cellcolor{red!10}{0.0} & \cellcolor{red!10}{\textbf{0.74}} & \cellcolor{red!10}{\textbf{0.09}} & \cellcolor{red!10}{0.01} & \cellcolor{red!10}{\textbf{0.0}} \\*
        (IA)\textsuperscript{3} & 229,376 & \cellcolor{red!10}{0.02} & \cellcolor{red!10}{\textbf{0.26}} & \cellcolor{red!10}{0.0} & \cellcolor{red!10}{0.0} & \cellcolor{red!10}{0.0} & \cellcolor{red!10}{0.0} & \cellcolor{red!10}{0.0} & \cellcolor{red!10}{0.0} & \cellcolor{red!10}{\textbf{0.0}} \\*
        Prompt tuning & 40,960 & 0.7 & \textbf{0.26} & 0.0 & 0.0 & 0.0 & 0.0 & \cellcolor{red!10}{0.08} & \cellcolor{red!10}{\textbf{0.25}} & \cellcolor{red!10}{\textbf{0.0}} \\

        \multicolumn{11}{l}{\cellcolor{gray!10}{\textbf{StarCoder2-3B}}} \bigstrut \\*
        None & 0 & 0.93 & 0.1 & 0.53 & 0.14 & 0.06 & 0.52 & \textbf{1.0} & 0.41 & 0.22 \\*
        Fine-tuning & 3,030,371,328 & 0.96 & 0.23 & (0.17) & \textbf{0.62} & \textbf{0.22} & (0.38) & \textbf{1.0} & 0.41 & (0.19) \\*
        LoRA & 4,546,560 & \textbf{0.98} & \textbf{0.24} & (0.19) & 0.59 & 0.2 & \textbf{0.53} & \textbf{1.0} & (0.32) & (0.16) \\*
        (IA)\textsuperscript{3} & 468,480 & 0.94 & 0.21 & (0.27) & 0.41 & 0.14 & (0.39) & \textbf{1.0} & \textbf{0.42} & \textbf{0.24} \\*
        Prompt tuning & 61,440 & 0.93 & 0.1 & \textbf{0.54} & 0.15 & 0.06 & (0.28) & \textbf{1.0} & (0.38) & (0.18) \\

        \multicolumn{11}{l}{\cellcolor{gray!10}{\textbf{Qwen2.5-Coder-3B}}} \bigstrut \\*
        None & 0 & 0.97 & 0.21 & 0.18 & 0.53 & 0.19 & 0.0 & \textbf{1.0} & 0.42 & 0.34 \\*
        Fine-tuning & 3,085,938,688 & \textbf{0.98} & \textbf{0.25} & (0.15) & 0.53 & (0.13) & 0.0 & \textbf{1.0} & (0.39) & (0.21) \\*
        LoRA & 3,686,400 & \textbf{0.98} & 0.23 & \textbf{0.2} & 0.55 & 0.19 & 0.51 & \textbf{1.0} & 0.45 & \textbf{0.4} \\*
        (IA)\textsuperscript{3} & 479,232 & 0.97 & 0.22 & 0.19 & \textbf{0.6} & \textbf{0.22} & \textbf{0.64} & \textbf{1.0} & \textbf{0.46} & 0.38 \\*
        Prompt tuning & 40,960 & 0.97 & 0.22 & \textbf{0.2} & 0.53 & 0.2 & 0.55 & (0.99) & (0.36) & (0.31) \\

        \multicolumn{11}{l}{\cellcolor{gray!10}{\textbf{CodeGen2-3.7B}}} \bigstrut \\*
        None & 0 & \cellcolor{red!10}{0.0} & \cellcolor{red!10}{0.0} & \cellcolor{red!10}{0.0} & \cellcolor{red!10}{0.0} & \cellcolor{red!10}{0.0} & \cellcolor{red!10}{0.0} & \cellcolor{red!10}{0.0} & \cellcolor{red!10}{0.0} & \cellcolor{red!10}{\textbf{0.0}} \\*
        Fine-tuning & 3,641,174,016 & \cellcolor{red!10}{0.41} & \cellcolor{red!10}{0.07} & \cellcolor{red!10}{0.05} & \cellcolor{red!10}{\textbf{0.33}} & \cellcolor{red!10}{\textbf{0.15}} & \cellcolor{red!10}{0.0} & \textbf{0.74} & \textbf{0.26} & \textbf{0.0} \\*
        LoRA & 4,194,304 & \cellcolor{red!10}{\textbf{0.42}} & \cellcolor{red!10}{0.07} & \cellcolor{red!10}{\textbf{0.12}} & \cellcolor{red!10}{0.09} & \cellcolor{red!10}{0.06} & \cellcolor{red!10}{\textbf{0.58}} & \cellcolor{red!10}{0.4} & \cellcolor{red!10}{0.04} & \cellcolor{red!10}{\textbf{0.0}} \\*
        (IA)\textsuperscript{3} & 458,752 & \cellcolor{red!10}{0.0} & \cellcolor{red!10}{0.0} & \cellcolor{red!10}{0.0} & \cellcolor{red!10}{0.0} & \cellcolor{red!10}{0.0} & \cellcolor{red!10}{0.0} & \cellcolor{red!10}{0.0} & \cellcolor{red!10}{0.0} & \cellcolor{red!10}{\textbf{0.0}} \\*
        Prompt tuning & 81,920 & \cellcolor{red!10}{0.22} & \cellcolor{red!10}{\textbf{0.26}} & \cellcolor{red!10}{0.0} & \cellcolor{red!10}{0.0} & \cellcolor{red!10}{0.0} & \cellcolor{red!10}{0.0} & \cellcolor{red!10}{0.0} & \cellcolor{red!10}{0.0} & \cellcolor{red!10}{\textbf{0.0}} \\

        \multicolumn{11}{l}{\cellcolor{gray!10}{\textbf{CodeLlama-7B}}} \bigstrut \\*
        None & 0 & \textbf{0.98} & 0.24 & 0.18 & 0.6 & 0.23 & 0.14 & 0.99 & 0.4 & 0.21 \\*
        Fine-tuning & 6,738,546,688 & \textbf{0.98} & 0.24 & \textbf{0.22} & \textbf{0.62} & (0.22) & (0.0) & \textbf{1.0} & \textbf{0.41} & \textbf{0.22} \\*
        LoRA & 8,388,608 & \textbf{0.98} & \textbf{0.26} & \textbf{0.22} & 0.61 & \textbf{0.24} & \textbf{0.51} & 0.99 & (0.35) & (0.2) \\*
        (IA)\textsuperscript{3} & 614,400 & \textbf{0.98} & 0.24 & 0.2 & (0.59) & \textbf{0.24} & 0.36 & \textbf{1.0} & (0.39) & 0.21 \\*
        Prompt tuning & 81,920 & (0.97) & (0.23) & (0.11) & (0.59) & (0.13) & 0.19 & 0.99 & (0.37) & (0.2) \\

        \multicolumn{11}{l}{\cellcolor{gray!10}{\textbf{CodeGen2-7B}}} \bigstrut \\*
        None & 0 & \textbf{0.99} & 0.22 & 0.18 & 0.57 & 0.21 & 0.35 & \textbf{1.0} & 0.38 & \textbf{0.13} \\*
        Fine-tuning & 6,862,858,240 & (0.98) & 0.23 & (0.15) & (0.51) & (0.2) & (0.27) & \textbf{1.0} & (0.35) & (0.08) \\*
        LoRA & 8,388,608 & \textbf{0.99} & \textbf{0.24} & (0.17) & \textbf{0.6} & 0.21 & 0.41 & \textbf{1.0} & (0.37) & (0.12) \\*
        (IA)\textsuperscript{3} & 917,504 & (0.98) & 0.22 & 0.18 & 0.58 & \textbf{0.22} & \textbf{0.52} & \textbf{1.0} & \textbf{0.39} & \textbf{0.13} \\*
        Prompt tuning & 81,920 & (0.98) & 0.22 & \textbf{0.19} & (0.54) & 0.21 & 0.51 & \textbf{1.0} & (0.36) & (0.12) \\

        \multicolumn{11}{l}{\cellcolor{gray!10}{\textbf{StarCoder2-7B}}} \bigstrut \\*
        None & 0 & 0.92 & 0.11 & \textbf{0.53} & 0.12 & 0.05 & 0.37 & \textbf{1.0} & 0.35 & 0.23 \\*
        Fine-tuning & 7,173,923,840 & \textbf{0.97} & 0.24 & (0.2) & 0.6 & 0.2 & (0.14) & \textbf{1.0} & \textbf{0.43} & (0.22) \\*
        LoRA & 7,340,032 & \textbf{0.97} & 0.24 & (0.21) & 0.56 & 0.2 & \textbf{0.53} & \textbf{1.0} & 0.42 & (0.22) \\*
        (IA)\textsuperscript{3} & 753,664 & 0.95 & 0.24 & (0.2) & \textbf{0.62} & \textbf{0.22} & 0.41 & \textbf{1.0} & 0.42 & (0.21) \\*
        Prompt tuning & 92,160 & 0.93 & \textbf{0.25} & (0.2) & 0.54 & 0.19 & 0.43 & \textbf{1.0} & 0.42 & \textbf{0.24} \\

        \multicolumn{11}{l}{\cellcolor{gray!10}{\textbf{Qwen2.5-Coder-7B}}} \bigstrut \\*
        None & 0 & 0.97 & 0.23 & 0.2 & 0.59 & 0.21 & 0.49 & \textbf{1.0} & 0.39 & 0.37 \\*
        Fine-tuning & 7,615,616,512 & (0.94) & (0.19) & (0.14) & (0.57) & (0.18) & (0.0) & \textbf{1.0} & \textbf{0.41} & (0.18) \\*
        LoRA & 5,046,272 & \textbf{0.98} & \textbf{0.25} & 0.23 & 0.62 & 0.21 & \textbf{0.51} & \textbf{1.0} & (0.36) & (0.32) \\*
        (IA)\textsuperscript{3} & 645,120 & \textbf{0.98} & 0.24 & \textbf{0.24} & \textbf{0.64} & 0.21 & (0.39) & \textbf{1.0} & 0.39 & 0.38 \\*
        Prompt tuning & 71,680 & (0.96) & 0.23 & 0.21 & 0.63 & \textbf{0.22} & (0.36) & \textbf{1.0} & \textbf{0.41} & \textbf{0.41} \\

        \multicolumn{11}{l}{\cellcolor{gray!10}{\textbf{Qwen2.5-Coder-14B}}} \bigstrut \\*
        None & 0 & \textbf{0.98} & 0.25 & \textbf{0.24} & 0.58 & 0.2 & 0.4 & \textbf{1.0} & 0.37 & 0.3 \\*
        Fine-tuning & 14,770,033,664 & (0.96) & (0.24) & (0.2) & (0.55) & (0.17) & (0.12) & \textbf{1.0} & \textbf{0.43} & 0.35 \\*
        LoRA & 12,502,912 & (0.97) & \textbf{0.26} & \textbf{0.24} & \textbf{0.61} & \textbf{0.21} & 0.53 & \textbf{1.0} & 0.41 & \textbf{0.38} \\*
        (IA)\textsuperscript{3} & 958,464 & \textbf{0.98} & \textbf{0.26} & \textbf{0.24} & \textbf{0.61} & 0.2 & (0.27) & \textbf{1.0} & 0.39 & 0.34 \\*
        Prompt tuning & 102,400 & \textbf{0.98} & 0.25 & (0.2) & 0.59 & 0.2 & \textbf{0.54} & \textbf{1.0} & 0.39 & \textbf{0.38} \\

        \multicolumn{11}{l}{\cellcolor{gray!10}{\textbf{StarCoderBase}}} \bigstrut \\*
        None & 0 & 0.92 & 0.11 & \textbf{0.48} & 0.15 & 0.06 & 0.28 & 0.99 & \textbf{0.4} & 0.2 \\*
        Fine-tuning & 15,517,456,384 & \textbf{0.97} & \textbf{0.26} & (0.22) & \textbf{0.58} & \textbf{0.22} & 0.41 & \textbf{1.0} & (0.39) & \textbf{0.22} \\*
        LoRA & 8,028,160 & 0.92 & 0.11 & \textbf{0.48} & 0.15 & 0.06 & \textbf{0.47} & 0.99 & (0.37) & 0.2 \\*
        (IA)\textsuperscript{3} & 1,239,040 & 0.92 & 0.11 & \textbf{0.48} & 0.16 & 0.06 & 0.46 & 0.99 & \textbf{0.4} & 0.2 \\*
        Prompt tuning & 122,880 & (0.91) & 0.13 & (0.34) & 0.2 & 0.09 & (0.25) & (0.78) & (0.22) & (0.1) \\

        \multicolumn{11}{l}{\cellcolor{gray!10}{\textbf{StarCoder2-15B}}} \bigstrut \\*
        None & 0 & 0.93 & 0.13 & \textbf{0.53} & 0.17 & 0.07 & 0.37 & \textbf{1.0} & 0.34 & 0.25 \\*
        Fine-tuning & 15,655,899,136 & \textbf{0.98} & 0.25 & (0.19) & 0.49 & 0.17 & (0.03) & (0.99) & 0.4 & (0.23) \\*
        LoRA & 12,124,160 & \textbf{0.98} & \textbf{0.26} & (0.21) & 0.61 & 0.22 & (0.16) & \textbf{1.0} & 0.38 & 0.26 \\*
        (IA)\textsuperscript{3} & 1,249,280 & 0.93 & 0.13 & \textbf{0.53} & 0.17 & 0.07 & 0.43 & \textbf{1.0} & 0.34 & 0.25 \\*
        Prompt tuning & 122,880 & \textbf{0.98} & \textbf{0.26} & (0.18) & \textbf{0.62} & \textbf{0.24} & \textbf{0.52} & \textbf{1.0} & \textbf{0.44} & \textbf{0.28} \\

        \multicolumn{11}{l}{\cellcolor{gray!10}{\textbf{CodeGen2-16B}}} \bigstrut \\*
        None & 0 & 0.98 & 0.23 & 0.18 & 0.57 & 0.22 & \textbf{0.7} & \textbf{1.0} & \textbf{0.39} & 0.13 \\*
        Fine-tuning & 16,032,155,648 & \textbf{0.99} & 0.24 & (0.13) & (0.51) & (0.15) & (0.05) & (0.98) & (0.28) & (0.07) \\*
        LoRA & 13,369,344 & \textbf{0.99} & \textbf{0.25} & (0.17) & (0.52) & (0.19) & (0.29) & \textbf{1.0} & (0.37) & \textbf{0.14} \\*
        (IA)\textsuperscript{3} & 1,462,272 & 0.98 & 0.23 & \textbf{0.19} & 0.57 & (0.21) & (0.53) & \textbf{1.0} & \textbf{0.39} & 0.13 \\*
        Prompt tuning & 122,880 & 0.98 & 0.23 & \textbf{0.19} & \textbf{0.61} & \textbf{0.24} & (0.54) & \textbf{1.0} & (0.37) & 0.13

    \end{xltabular}
\end{ThreePartTable}

\subsubsection{Syntactic Correctness}
\label{sec:syntax}
As shown in \Cref{tab:eval-summary}(b), the percentage of syntactically correctly generated unit tests is relatively high across the models, with $>$ 91\% for all except the smaller CodeGen2-1B and CodeGen2-3.7B models (marked in red in \Cref{tab:eval-summary}). These two cannot produce any valid unit tests with the base model. However, after tuning, both models are able to produce up to 75\% and 42\% syntactically valid tests, respectively. Notable findings from \Cref{tab:eval-summary}(b)) are as follows: 1) The increase of 70\% using prompt tuning of CodeGen2-1B, tuning only 0.02‰ (40 thousand out of 1 billion parameters) of the parameters. 2) The increase of 42\% using LoRA on CodeGen2-3.7B, tuning only 0.1\% (4.19 million out of 3.7 billion parameters) of the parameters.

Further, data in \Cref{tab:eval-summary}(b) illustrate that: 1) Full fine-tuning achieves higher percentages of syntactically correct unit tests than the baseline. The median value of syntactically valid tests for the full fine-tuning is 0.97, which is 1\% higher than the value of the baseline (0.96). 2) There is a slight tendency for models with more trainable parameters to generate a higher proportion of syntactically correct unit tests. 3) The difference in syntactical validity between fine-tuning and PEFTs is not very prominent. The median values of syntactically valid tests of PEFT methods are: LoRA 0.98, (IA)\textsuperscript{3} 0.95, and prompt tuning 0.96. However, considering that full fine-tuning often tunes parameters several thousand more times than PEFT methods, it is probably not wise to use full fine-tuning to improve the syntactic correctness of the unit tests if cost is a concern.
   
\subsubsection{Similarity}
The \textsc{Methods2Test\textsubscript{runnable}} dataset CodeBLEU column in \Cref{tab:eval-summary}(b) shows the CodeBLEU scores of the baseline, full fine-tuning, and PEFT methods. Due to the low number of syntactically valid tests generated from the CodeGen2-1B and CodeGen2-3.7B models, we exclude these models from further analysis. The median CodeBLEU scores are: baseline 0.21, fine-tuning 0.24, LoRA 0.24, (IA)\textsuperscript{3} 0.22, and prompt tuning 0.23.
The notable results are: 1) LoRA achieve the highest CodeBLEU scores in seven of the eleven valid models, improving the median CodeBLEU value by 14\% over the baseline. 2) Prompt tuning is also the only PEFT method that shows slight negative results.

\subsubsection{Passing rate}
\begin{table}[htb]
    \newcolumntype{Y}{>{\centering\arraybackslash}X}
    \centering
    \caption{Test execution statuses for StarCode2-7B.}
    \label{tab:test-statuses}
    \small
    \begin{tabularx}{\columnwidth}{lYYYYYY}
    \toprule
    \textbf{Method} & \textbf{Succ.} & \textbf{Failed} & \textbf{Interrupt} & \textbf{CompErr} & \textbf{NoAssert} \\
    \midrule
    None & 53\% & 3\% & 3\% & 16\% & 26\% \\
    Fine-tuning & 20\% & 14\% & 9\% & 51\% & 6\% \\
    LoRA & 21\% & 15\% & 9\% & 51\% & 5\% \\
    (IA)\textsuperscript{3} & 20\% & 14\% & 10\% & 49\% & 7\% \\
    Prompt tuning & 20\% & 15\% & 7\% & 52\% & 6\% \\
    \bottomrule
    \end{tabularx}
\end{table}
As explained in \cref{sec:test-execution}, we consider a test as passing if it executes successfully without errors, as determined by a test oracle. 
Any tests that did not run due to an environmental exception (see \cref{sec:test-execution}) are excluded from analysis because such cases did not reflect the quality of the generated test cases. 
As we use \textsc{Methods2Test\textsubscript{runnable}} to generate and evaluate tests, the original focal method under test serves as a proxy for the oracle, effectively acting as the ground truth. While we do not assume that the methods under test are bug-free, we trust that all methods from the \textsc{Methods2Test} dataset are at least passable and thus suitable as oracles for evaluating generated tests. However, using the method as the oracle of test passing introduces a limitation: tests performing no assertion checks may still return passing results, which are misleading. To address this issue, we heuristically mark the generated tests as failed (even if their executions return passing results) if they do not contain any of the substrings ``\texttt{assert}'', ``\texttt{verify}'', or ``\texttt{fail}'', which are indicative of meaningful assertions or validation logic in the test body.

From the pass@1 column in \Cref{tab:eval-summary}(b), we calculated individual median pass@1 values and got the results: baseline 0.20, fine-tuning 0.17, LoRA 0.21, (IA)\textsuperscript{3} 0.20, and prompt tuning 0.20. The baseline achieves the highest pass@1 rate on five of the eight models, with a significantly higher median score than all tuning methods. To understand this phenomenon, we analyzed and categorized different reasons for tests not returning a passing result. \Cref{tab:test-statuses} shows an example of statistical analysis results of the execution status of test cases generated by the StarCode2-7B model. We found that the primary cause of the low number of successful test executions after fine-tuning is compilation errors. Although most of the generated content is syntactically valid (see \cref{sec:syntax}), compilation errors still happen, mainly due to the generated content calling nonexistent methods or having type mismatches. Most of these compilation errors stem from repetition or boilerplate patterns that superficially satisfy the heuristic of unit test generation without actually validating the method’s behavior.

\subsubsection{Coverage}
Among the tests returning passing results, the individual median values of the instruction coverage are: baseline 0.53, fine-tuning 0.55, LoRA 0.59, (IA)\textsuperscript{3} 0.58, and prompt tuning 0.54. The individual median values of the branch coverage are: baseline 0.19, fine-tuning 0.18, LoRA 0.20, (IA)\textsuperscript{3} 0.21, and prompt tuning 0.20. These numbers show that the code coverage is generally high among the tuned models. It means that if the tuned models can create runnable tests, the tests are likely to cover a significant portion of the code to potentially identify edge cases. 
The results also show that LoRA and full fine-tuning outperform other methods, although the differences between their performance and other PEFT methods are insignificant. 

\subsubsection{Mutation score}
\label{sec:mutation-score}
Of the tests that pass and returning passing results, the individual median values of the mutation scores are: baseline 0.37, fine-tuning 0.12, LoRA 0.51, (IA)\textsuperscript{3} 0.43, and prompt tuning 0.44. A notable finding is that full fine-tuning underperforms the baseline, suggesting that extensive parameter updates may diminish the model’s ability to generate fault-revealing tests. Conversely, all PEFT methods outperform the baseline. LoRA is the most performant PEFT method in five of the eleven models, improving the median mutation scores value by 38\% over the baseline.

\subsubsection{Resilience against catastrophic forgetting}
\label{sec:catastrophic-forgetting}
The results in \Cref{tab:eval-summary}(c) show the performance of the baseline and tuned models on a normal code generation task using the Java subset of the HumanEval-X dataset. The results show some cases of degradation (marked using parentheses in \Cref{tab:eval-summary}(c)) of code generation when comparing the results of the tuned models with the baseline model. This applies both to full fine-tuning and PEFT. However, the levels of degradation are, for the most part, trivial and inconsistent. The median CodeBLEU scores are: baseline 0.39, fine-tuning 0.40, LoRA: 0.37, (IA)\textsuperscript{3} 0.39, and prompt tuning 0.37. The median pass@1 values are: baseline 0.22, fine-tuning 0.21, LoRA 0.20, (IA)\textsuperscript{3} 0.21, and prompt tuning 0.20. In several cases, e.g. StarCoder2-15B, PEFT tuning actually improves the CodeBLEU scores and pass@1 value. This shows that tuning methods are mostly resilient against catastrophic forgetting, and PEFT methods exhibit transfer learning capabilities comparable to those achieved through full fine-tuning.

\vspace{1em}
\noindent\fbox{%
    \parbox{\columnwidth}{%
        \textbf{Summary for RQ1}: PEFT could be a better choice than full fine-tuning if cost-effectiveness to improve the syntactic correctness of the generated unit tests is a concern. LoRA can generate test cases as good as full fine-tuning, with consistently better mutation score. The tuned models may generate fewer executable test cases than baseline models, although the executable ones are better at test coverage. Further, PEFT methods are mostly resilient against catastrophic forgetting.
    }%
}

\subsection{RQ2: PEFT training resource utilization effectiveness}
One of the main advantages of using PEFT methods is that they require significantly less computing during training, mainly because of reduced backpropagation calculations. The fewer trainable parameters, the more efficient. As seen from the second column in \Cref{tab:eval-summary}(a), PEFT methods show a significant reduction of trainable parameters compared to full fine-tuning. This is also essential to fit big models into GPU memory. For example, we could train the largest 16 billion parameter model using a single NVIDIA A100 80GB GPU for LoRA, whereas we needed 6 x NVIDIA A100 40GB GPUs for full fine-tuning.

\begin{figure}[htbp]
    \centering
    \includegraphics{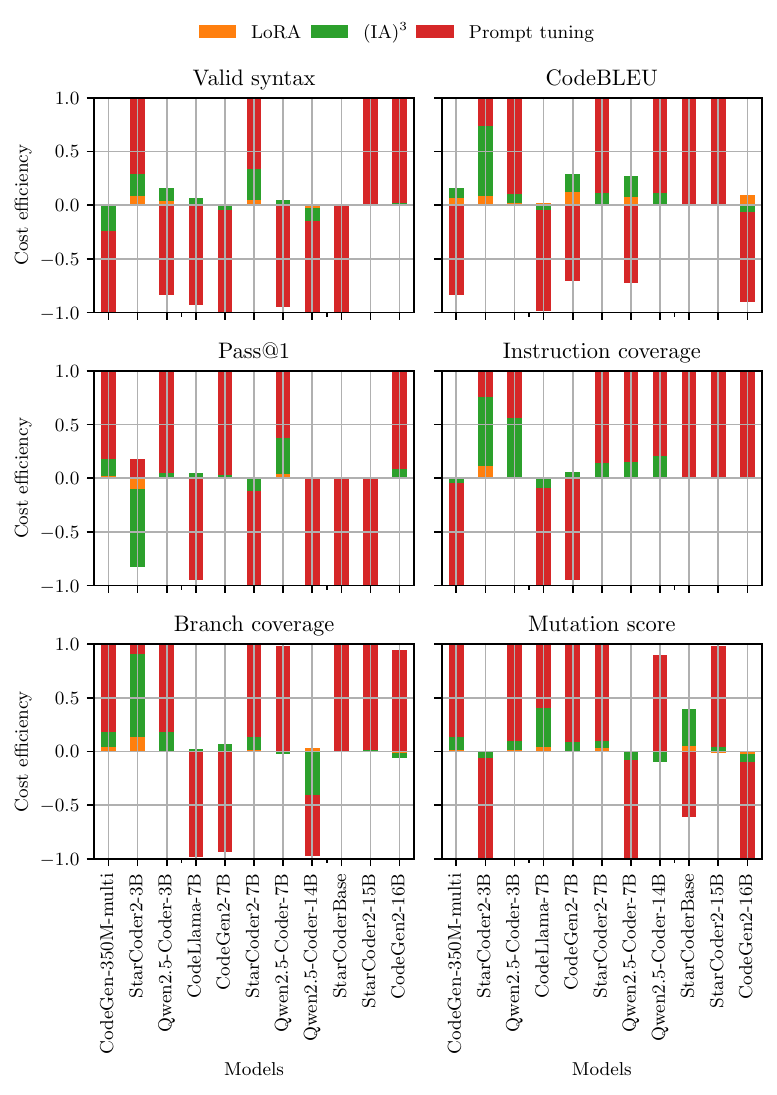}
    \caption{Stacked bar chart of symmetrically normalized metrics value differences (increase or decrease) between baseline and tuned models, divided by the percentage of trainable parameters of models, on the \textsc{Methods2Test\textsubscript{runnable}} dataset.}
    \label{fig:relative-importance}
\end{figure}

\Cref{fig:relative-importance} shows the results of the metrics (valid syntax, CodeBLEU, pass@1, instruction coverage, branch coverage, and mutation score) for each tuning method relative to the percentage of trainable parameters, symmetrically normalized between -1 and 1. As shown in \Cref{fig:relative-importance}, prompt tuning has the highest efficiency for every metric and is the method that you get the ``most bang for your buck.'' Moreover, prompt tuning has the most potential cost-effectiveness on models with large sizes, i.e., Qwen2.5-Coder-14B, StarCoderBase, StarCoder2-15B, and CodeGen2-16B. This can be explained by the increased capacity of larger models to better understand and leverage the additional context provided by prompt tuning. However, prompt tuning does not show consistent superiority for all the models. The method contributes to negative performance in three models (StarCoder2-3B, CodeLlama-7B, and CodeGen2-7B). 

For the four smallest models (Qwen2.5-Coder-3B, CodeGen-350M-multi, CodeGen2-1B, and StarCoder2-3B), (IA)\textsuperscript{3} clocks in at second place, which provides better results than LoRA. However, (IA)\textsuperscript{3} generally fails to efficiently improve the three largest models. LoRA achieves a relatively low efficiency compared to the other PEFT methods. However, with the exception of pass@1, it does not lead to any significant negative impact on the performance of any of the models. Full fine-tuning obviously has the worst efficiency, as it needs to train all the model parameters.

\vspace{1em}
\noindent\fbox{%
    \parbox{\columnwidth}{%
        \textbf{Summary for RQ2}: PEFT methods offer significantly reduced computational requirements compared to full fine-tuning and are ranked by resource utilization effectiveness as follows: prompt tuning $>$ (IA)\textsuperscript{3} $>$ LoRA $>$ full fine-tuning. Prompt tuning demonstrates the highest returns on investment for most models but also carries risks of negative performance impact. 
    }%
}

\section{Discussion}
\label{chap:discussion}
This section highlights our novel insights by comparing our results with related work. We also discuss possible threats to the validity of our study and how we mitigate them. 

\subsection{Comparison with related studies}
As the effectiveness of PEFT methods is task-dependent, our results provide novel insights into their performance in the context of unit test generation.  \citet{Ayupov2022ParameterEfficientFO} conclude that LoRA is less effective than full fine-tuning for code generation. We show contradictory results that LoRA could be as good as or even more effective than fine-tuning in unit test generation. \citet{weyssow2024exploring} explored generating code from natural language instructions and concluded that LoRA is the most effective among the PEFT methods they evaluated. Our results show that prompt tuning is the most effective in terms of cost and resource utilization in unit test generation, although LoRA provides better performance without considering the cost aspects. 

Compared to the existing studies, we also performed a more comprehensive evaluation, as we invested significant efforts to compare all PEFT methods to full fine-tuning across all model sizes (up to 16 billion parameters). In contrast, \citet{Ding2023ParameterefficientFO} only evaluated PEFT using the 355 million parameter RoBERTa\textsubscript{LARGE} model, while \citet{weyssow2024exploring} only did full fine-tuning for smaller models up to a maximum of 350 million parameters. While existing works on code generation or test generation have demonstrated promising results, they are often limited to artificially constructed benchmarks such as HumanEval. These datasets are typically small, synthetically curated, and lack the complexity of real-world software systems. As our results show, such controlled settings also overlook critical challenges that arise in real-world scenarios, such as missing dependencies, inconsistent, or erroneous code. Defects4J \citep{just2014defects4j}, one of the more realistic benchmarks available, contains only 17 repositories. This starkly contrasts the scale and messiness of actual industrial or open-source projects. As a result, prior work such as \citet{alagarsamy2024a3test} may significantly overestimate the robustness and generalizability of models when deployed beyond synthetic testbeds. Our evaluation, grounded in more realistic settings using \textsc{Methods2Test} \citep{tufano2021unit}, highlights these limitations and emphasizes the need for evaluation frameworks that reflect the true complexity of software development.

Several studies have leveraged LLMs for generating unit tests \citep{tufano2021unit, siddiq2024using, yuan2023manual, schäfer2023empirical, steenhoek2023reinforcement}. However, none of the works investigate the potential performance of using PEFT compared to full fine-tuning. \citet{tufano2021unit} automatically generated unit tests and reported 95\% syntactic correctness when training BART large (400 million parameters) \citep{lewis2019bart} on \textsc{Methods2Test}. With PEFT methods, we can generate an average median of 96\% syntactically correct unit tests. Training between 20 thousand and 13 million parameters on 1\% of the same data, showcasing the potentially powerful performance of PEFT methods.

Different from studies \citep{siddiq2024using, yuan2023manual, schäfer2023empirical} using large instruction-tuned enterprise LLMs such as ChatGPT to generate test cases, this study focused on generating test cases from autoregressive base code models. Compared to e.g. \citet{schäfer2023empirical}, our best model produced 13\% less statement coverage and 55\% lower branch coverage. Achieving state-of-the-art performance was not the focus of our work. Instead, our work provides an empirical evaluation of PEFT methods in a more practical scenario, where organizations can efficiently customize models for their own testing needs. One reason for the reduced coverage is probably due to the size differences of the models. The largest model we evaluated had 16 billion paramaters, opposed to the lager commercial GPT-3.5-Turbo. Our results shown in \Cref{tab:eval-summary} illustrate that larger models generally generate better unit tests than smaller ones. Another reason is probabaly due to the differences between the benchmark datasets used by our study. We use a real-world evaluation dataset that is more diverse, spanning 119 repositories, versus the 25 npm packages used by \citet{siddiq2024using}. Our analyses on the reasons for low pass@1 rates show that it is necessary to consider potential compilation and code dependency issues when using LLMs to generate executable unit tests, providing valuable novel insights to improve unit test generation using PEFT or ChatGPT. Opposed to related works like \citep{siddiq2024using, yuan2023manual, schäfer2023empirical}, we conduct mutation testing. As our results depict in \cref{sec:mutation-score}, measuring mutation score is crucial to ensure generated tests are actually able to reveal faults.

\subsection{Implications to Academia and Industry}
Writing unit tests is a hard and often time-consuming task \citep{7102609}. However, they are essential to ensure code quality, stability, and maintainability \citep{Gonzalez2017ALS}. Automating unit test generation using LLMs can significantly reduce the time and effort required for this process, making it an attractive solution for academia and industry. As our experiments show, full fine-tuning is sometimes required to achieve adequate quality of the generated unit tests, which is computationally demanding. Many LLM users may not have the necessary computing resources to fine-tune LLMs fully. Our results show that various PEFT methods can achieve similar performance to full fine-tuning by training a minuscule parameter fraction. Furthermore, our results suggest starting with prompt tuning for cost-effectiveness. Then, if stability issues arise, LoRA should be considered as a reliable alternative because it does not lead to a negative performance impact. 

\subsection{Threats to Validity}

\subsubsection{External validity}
One of the threats to external validity is the selection of models. We minimize this threat by selecting a broad range of models varying in size, pre-training data, and learning objectives. As we have have spent upward of 700 GPU hours for model tuning alone, exploring an even broader range of alternative models is constrained due to cost. Another potential concern is the validity of the benchmarking dataset. We chose to use \textsc{Methods2Test}, which is the largest publicly available corpus of real-world Java unit test cases and corresponding focal methods. We believe this provides a representative evaluation setting compared to existing studies like \citet{alagarsamy2024a3test}, although future work should expand the analysis to other programming languages to ensure generalizability. The last threat is related to our evaluation metric. While popularly used, we recognize BLEU as a sub-optimal performance estimator for code generation. We, therefore, employ a more code-friendly version, CodeBLEU. Although CodeBLEU may not be the optimal metric to evaluate the functional correctness of unit tests completely, it provides a good tradeoff between evaluation effort and validity. To supplement our evaluation, we have also spent significant effort developing evaluation pipelines, as well as over 200 CPU hours for executing the test in the \textsc{Methods2Test\textsubscript{runnable}} dataset, calculating pass@1, code coverage, and mutation scores.

\subsubsection{Internal validity}
The main threat to internal validity is comprised of the selection of hyperparameters. We use the hyperparameters reported in the introductory paper of each PEFT method to avoid introducing unnecessary biases. Additionally, we use greedy decoding to support reproducibility and reduce variability in the generated results.

\section{Conclusion and Future Work}
\label{chap:conclusion}
Although PEFT has been investigated to boost the quality and efficiency of code-related tasks using LLMs, they have not been studied in the context of unit test generation. We performed the first empirical study to compare different PEFT methods and full fine-tuning to generate unit tests. Our comparison involved thirteen LLMs of different sizes and architectures. Our results offer valuable insights into the advantages and challenges of using PEFT and full fine-tuning for unit test generation. With the knowledge we've gathered, individuals can choose the most suitable PEFT methods or full fine-tuning based on the size and characteristics of the LLMs. In future work, we plan to extend the analysis perform an empirical comparison of methods for tuning instruction-based LLMs, covering both PEFT methods per se and variations in prompt templates

In addition to unit tests, manually coding other automated test cases, such as integration and system tests, can be cost-intensive. In the future, we intend to explore the utilization of LLMs in conjunction with cost-effective tuning methods to automatically generate other types of tests.

\section*{Acknowledgements}
The empirical results were supported through the computational resources of HPC facilities at the Norwegian University of Science and Technology (NTNU) \citep{sjalander2019epic}.

\section*{Data Availability}
\noindent The \textsc{Methods2Test\textsubscript{small}} dataset is available at:\\
\url{https://huggingface.co/datasets/andstor/methods2test_small}\\

\noindent The \textsc{Methods2Test\textsubscript{runnable}} dataset is available at:\\
\url{https://huggingface.co/datasets/andstor/methods2test_runnable}\\

\noindent The \textsc{Methods2Test\textsubscript{meta}} dataset is available at:\\
\url{https://huggingface.co/datasets/andstor/methods2test_meta}\\

\noindent Metadata and links to the trained models can be found at:\\
\url{https://huggingface.co/datasets/andstor/peft-unit-test-generation-experiments}\\

\noindent Experiment code and results are available at:\\
\url{https://github.com/andstor/peft-unit-test-generation-replication-package}

\printbibliography

\end{document}